# Generation of high-quality terahertz OAM mode based on soft-aperture difference frequency generation

**Katsuhiko Miyamoto,**[1,2,*] **Kazuki Sano,**[1] **Takahiro Miyakawa,**[1] **Hiromasa Niinomi,**[3] **Kohei Toyoda,**[2] **Adam Vallés,**[2] **and Takashige Omatsu**[1,2]

[1] *Graduate School of Engineering, Chiba University, 1-33 Yayoi-cho, Inage-ku, Chiba 263-8522, Japan*
[2] *Molecular Chirality Research Center, Chiba University, 1-33 Yayoi-cho, Inage-ku, Chiba 263-8522, Japan*
[3] *Institute for Materials Research, Tohoku University, 2-1-1 Katahira Aoba-ku, Sendai 980-8577, Japan*
*\* k-miyamoto@faculty.chiba-u.jp*

**Abstract:** We demonstrate the generation of high-quality tunable terahertz (THz) vortices in an eigenmode by employing soft-aperture difference frequency generation of a vortex output and a Gaussian mode. The generated THz vortex output exhibits a high-quality orbital angular momentum (OAM) mode with a topological charge of $\ell_{THz} = \pm 1$ in a frequency range of 2-6 THz. The maximum average power of the THz vortex output obtained was ~3.3 µW at 4 THz.

## 1. Introduction

A terahertz (THz) wave with a low photon energy within the range of 1-30 meV enables the assignment of intermolecular vibrations or cluster-cluster interactions [1–7], and offers a variety of fundamental and applied sciences, so-called THz photonics, such as the morphological change of polymers [8], orientation control of orbital ordering planes in strongly correlated materials [9], and structural analysis of crystalline objects [10–14]. An optical vortex [15–20] possesses unique properties, such as an annular spatial profile and an orbital angular momentum (OAM), characterized by a topological charge $\ell$, associated with its helical wavefront. Thus, it has been widely studied in many applications, including optical trapping and manipulation [21–26], ultrahigh density telecommunications [27–31], nonlinear spectroscopy [32–34], the fabrication of chiral materials [35–42], and super-resolution microscopy [43–47]. Going beyond conventional THz photonics, THz vortex modes will open doors towards entirely advanced technologies, for instance, 2-dimensional THz imaging systems for structural identification of biomaterials with an ultrahigh spatial resolution beyond the diffraction limit.

To date, several artificially designed devices to generate THz vortex modes have been proposed, such as a molded phase plate composed of V-shaped slit antennas [48], a circular array antenna [49], an achromatic polarization device [50], and a binary diffractive element [51]. However, their conversion efficiencies from an incident Gaussian beam to an optical vortex mode are typically limited to a few percent owing to their narrow spectral bandwidth and low transmittance. Also, liquid crystal q-plates enable us to generate broadband THz vortices [52,53]. However, they exhibit relatively low damage threshold, and severe transmission loss. Furthermore, their bandwidths are not still enough to develop tunable THz vortex sources.

In recent years, we have developed a spiral phase plate (SPP) formed of a polymeric material, dubbed Tsurupica, with an extremely low-frequency dispersion ($dn/d\nu = -5.1 \times 10^{-4}$ /THz) and high transmission (0.4–0.94 cm$^{-1}$) in the THz frequency region (1–6 THz), and we have successfully generated monochromatic 2 and 4 THz vortex outputs with topological charges of

ℓ = 1 and 2 at high conversion efficiencies of over 70% and 50%, respectively [54]. The generation of a monocycle THz vortex mode [55] with ultrahigh intensity has also been demonstrated by employing the SPP in combination with tilted-pulse-front optical rectification [56]. Such a SPP is typically designed for a specific frequency; therefore, it inherently constrains the wavelength tunability of the THz vortex modes. This also leads to undesired higher-order radial Laguerre-Gaussian ($LG_{p,\ell}$) components [57], described by a radial index $p$, which thus degrades the mode quality of the generated vortex modes associated to the index $\ell$.

An alternative route to generate the THz vortex mode is to use optical vortex pumped difference frequency generation (DFG). We have successfully developed mid-infrared optical vortex sources based on 2 µm optical vortex laser[58]-pumped DFG with a moderate (millijoule level) pulse energy [58,59] and an extremely wide tunability over the entire mid-infrared region (6–18 µm) [60]. However, there is little discussion of the mode purity in an eigenmode of the vortex output generated from the DFG.

Here, we extend our previous mid-infrared optical vortex work to generate a tunable THz vortex mode with high quality in an eigenmode. This is the first demonstration of tunable THz vortex generation with a frequency range of 2–6 THz, based on the optical vortex-pumped DFG from a 4'-dimethylamino-N-methyl-4-stilbazolium tosylate (DAST) crystal. The generated vortex mode exhibits ultrahigh mode purity (estimated mode purity of >90%) by soft aperture effects for the removal of undesired higher-order radial modes due to the spatial overlap between the vortex and Gaussian pump beams in the soft-aperture DFG.

## 2. Experiments

### 2.1 Setup

Figure 1(a) shows the experimental setup of the tunable THz vortex source. An in-house-built picosecond laser (wavelength: 1064 nm, average output power: ~5 W, pulse width: 8.3 ps, pulse repetition frequency (PRF): 1 MHz) was used as a pump laser, and its output was delivered to two injection-seeded optical parametric amplifiers (OPA1 & 2) formed of two periodically poled stoichiometric lithium tantalite (PPSLT) crystals with a fan-out grating (15×1×35 mm, Λ = 29–31 µm) [61]. The OPA1 output was converted into a first-order optical vortex ($\ell = 1$) using an SPP (Fig. 1(b)), and the wavelength, $\lambda_1$, was then fixed to be 1.56 µm.

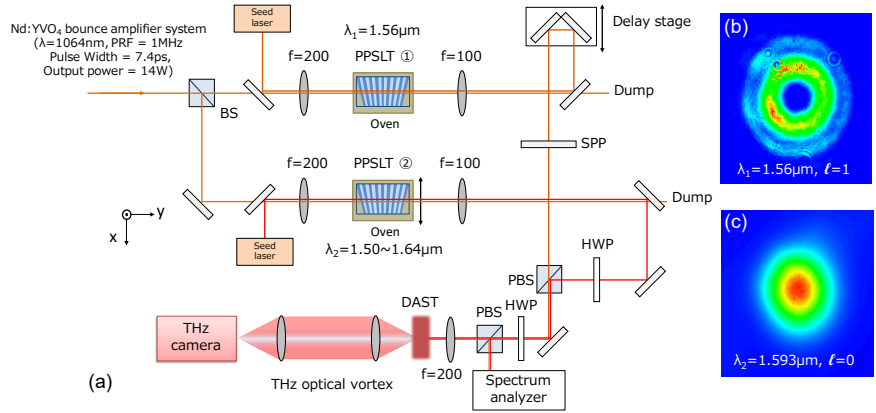

Fig. 1. (a) Experimental setup for tunable THz vortex generation based on 1.5 µm dual wavelength pumped DFG. (b) Spatial profile of OPA1 output with a wavelength $\lambda_1$, of 1.56 µm. (c) Spatial profile of OPA2 output with wavelength $\lambda_2$. $\lambda_2$ was tuned in the wavelength range of 1.50-1.64 µm.

The OPA2 output exhibited a Gaussian spatial form (Fig. 1(c)), and its wavelength, $\lambda_2$, was tuned within the wavelength range of 1.50–1.64 µm by translating the PPSLT crystal with a fan-out grating along the *x*-axis (as shown in Fig. 1) and controlling the lasing wavelength of the seed laser, an external cavity diode laser (ECDL; Anritsu, MG9541A). Both OPA1 and OPA2 outputs were then spatially overlapped and focused onto the DAST crystal, obtaining a ~170 µm diameter spot size thus facilitating the generation of a THz vortex beam by means of DFG type-0 geometry (all optical fields involved in the process have the same polarization). The frequency of the generated THz output with this system was seamlessly tuned within the frequency range of 2–6 THz. The pump beams used exhibited a linewidth of ~60 GHz, thus, the resulting THz vortex should exhibit a linewidth of less than double of ~ 120 GHz.

## 2.2 Results

A general relationship,

$$\ell_{THz} = (\ell_{OPA1} - \ell_{OPA2}) \frac{\lambda_{OPA2} - \lambda_{OPA1}}{|\lambda_{OPA2} - \lambda_{OPA1}|}, \quad (1)$$

is established, where $\ell_{THz}$, $\ell_{OPA1}$, and $\ell_{OPA2}$ are the topological charges of the THz, OPA1, and OPA2 outputs, according to the OAM conservation law [59]. It should be noted that the sign of the topological charge of the THz vortex is determined by the magnitude relation between the two pump wavelengths of the OPA1 and OPA2 outputs, as reported in our previous publications [59,60], *i.e.*, for $\lambda_1 < \lambda_2$ ($\lambda_1 > \lambda_2$), the THz vortex mode exhibits a positive (negative) $\ell_{THz}$, as depicted in Fig. 2(a). The generated positive vortex output with a positive $\ell_{THz}$ exhibited almost the same tuning curve as that for the negative vortex, due to the relatively wide acceptance bandwidth of the DAST crystal [62] (Fig. 2(b)). The vortex output was generated most efficiently at 4 THz, and its maximum average power was measured to be 3.3 µW at a pump power of 1 W, which corresponds to an optical-optical and a photon-photon conversion efficiency (from 1.5 µm dual-pump output to THz vortex output) of ~3.3×10$^{-4}$ and ~0.02 %, respectively (Fig. 2(c)). The conversion efficiency should be doubled by the refinements of coupling optics for pump beams. A key to improve further the conversion efficiency in our system is the power scaling of the pump source [54,63].

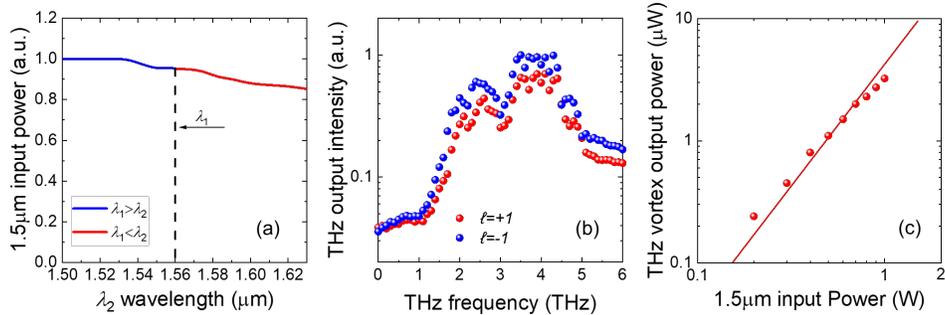

Fig. 2. (a) Tunability of Gaussian output from OPA2. The broken line shows the lasing wavelength of the vortex output from OPA1. (b) Frequency tunability of the THz vortex output from DAST-DFG. The red and blue circles represent experimental plots with a topological charge of $\ell_{THz} = \pm 1$. (c) Average power of the THz vortex at the 4 THz output as a function of the 1.5 µm dual-wavelength incident power.

The generated vortex mode exhibited a single ringed (*p*=0) intensity profile for both the near-field and far-field (focused plane), which indicates the high quality in an eigenmode, as shown in Fig. 3. The topological charge was also assigned to be ±1, as evidenced by the left or right diagonal Hermite-Gaussian mode HG$_{0,1}$, produced by the astigmatic focusing optics [64,65]. Such a high-quality vortex mode without undesired higher-order radial modes could be generated by the soft-aperture effects in the DFG (soft-aperture DFG). Figure 4 summarizes the spatial profiles of the focused vortex modes at various frequencies. Though it is difficult to identify the annular spatial form of 2 THz vortex output owing to the relatively low sensitivity of THz camera for 2 THz [66], the system does not impact the vortex output generation at a frequency below 2THz.

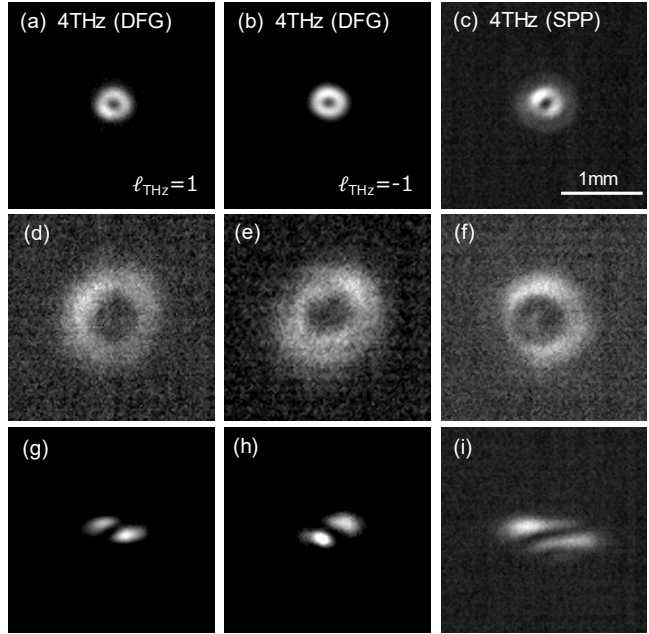

Fig. 3. Experimental (a,b) far- and (d,e) near-fields of the 4 THz vortex outputs with $\ell_{THz} = \pm 1$ generated from the present system. Experimental (c) far- and (f) near-fields of the 4 THz vortex output with $\ell_{THz} = 1$ produced with the Tsurupica-SPP. (g-i) Spatial forms of the focused 4 THz vortex outputs with an inclined lens.

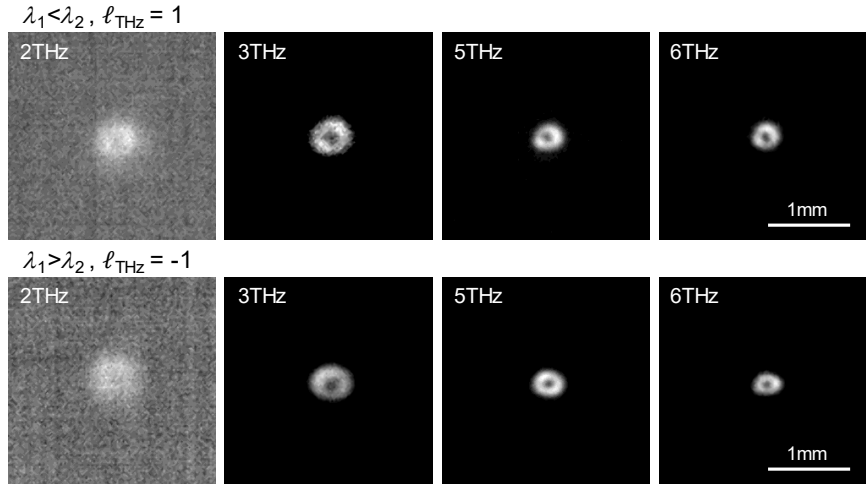

Fig. 4. Spatial forms of the THz vortex outputs with a topological charge of $\ell = \pm 1$ at frequencies of 2, 3, 5 and 6 THz.

## 3. Discussion

A conventional SPP provides only an azimuthal phase shift of $2\ell\pi$ to a fundamental Gaussian beam. Thus, the generated vortex output typically includes undesirable higher-order radial Laguerre-Gaussian (LG$_{p,\ell}$) modes with radial index $p$ and topological charge $\ell$ [57]. The electric field $u_{vortex}(r, \phi, \omega)$, of the vortex output with a beam radius of $\omega$ is then given by:

$$u_{vortex}(r,\phi,\omega) = c_0 e^{i\ell\phi} e^{-\frac{r^2}{\omega^2}} = \sum_p a_{p,\ell} \cdot u_{p,\ell}^{LG}(r,\phi,\omega), \quad (2)$$

$$u_{p,\ell}^{LG}(r,\phi,\omega) = c_{p,\ell} L_p^{|\ell|}\left(\frac{2r^2}{\omega^2}\right)\left(\frac{\sqrt{2}r}{\omega}\right)^{|\ell|} e^{i\ell\phi} e^{-\frac{r^2}{\omega^2}}, \quad (3)$$

where $r$ and $\phi$ are the radial and azimuthal coordinates, $\omega$ is the beam radius, $\ell$ is the topological charge of the SPP, $u_{p,\ell}^{LG}(r,\phi,\omega)$ is the LG mode with two mode indices $(p, \ell)$, $c_0$ and $c_{p,\ell}$ are the normalization constants, $a_{p,\ell}$ is the relative amplitude of the LG$_{p,\ell}$ mode, and $L_p^{|\ell|}$ is the generalized Laguerre polynomial. The relative intensity $\eta_{p,\ell}$, of the LG$_{p,\ell}$ mode is written as:

$$\eta_{p,\ell} = |a_{p,\ell}|^2 = \left|\iint u_{p,\ell}^{LG*} \cdot u_{vortex} r dr d\phi\right|^2, \quad (4).$$

When $\ell = 1$ or 2, the resulting $\eta_{p=0,\ell}$, defined as the purity of the vortex output in the eigenmode, is estimated to be 78.5% or 50%, respectively, which is typically the highest purity of the vortex output generated using the SPP.

In the present system, a vortex output with a wavelength of $\lambda_1$ (OPA1 output) and a Gaussian beam with a wavelength of $\lambda_2$ (OPA2 output) are focused and spatially overlapped on a nonlinear crystal, *i.e.* far-field, in which the individual radial modes with multiple rings ($p+1$) in the vortex output (OPA1 output) are spatially separated, to generate the THz vortex mode via DFG (Fig. 5(a)). The resulting DFG vortex output is expressed as:

$$u_{DFG}(r,\phi,\omega_1,\omega_2) \propto \left(\sum_p a_{p,\ell} \cdot u_{p,\ell}^{LG}(r,\phi,\omega_1) \cdot e^{i(2p+\ell)\frac{\pi}{2}}\right) \cdot e^{-\frac{r^2}{\omega_2^2}}, \quad (5)$$

where $\omega_1$ and $\omega_2$ are the beam radii of the vortex and Gaussian modes in the far-field, which indicate that the focused Gaussian mode for DFG with a nonlinear crystal plays a role as a soft-aperture to remove undesired higher-order radial LG modes (Fig. 5(b)). The resulting purity $\eta_{0,\ell}^{DFG}$ of the generated vortex mode is described as:

$$\eta_{0,\ell}^{DFG}(\omega_1,\omega_2) = \left|\iint u_{0,\ell}^{LG}(r,\phi,\omega_{DFG})^* \cdot u_{DFG}(r,\phi,\omega_1,\omega_2) r dr d\phi\right|^2, \quad (6)$$

where $\omega_{DFG}$ ($=\omega_1\omega_2/(\omega_1^2+\omega_2^2)^{1/2}$) is the beam radius of the difference frequency output, i.e., the THz vortex output, in the far-field. The $\eta_{0,1}^{DFG}$ ($= 98\%$) is estimated to be far beyond 78.5% (the purity of the first-order vortex mode generated by the SPP) at $\omega_1 = \omega_2$.

Figure 6 (a) shows the simulated spatial profile of the focused vortex mode generated by the SPP. The individual high-order radial LG modes included in the optical vortex are spatially separated, thereby forming a thick ring. They are removed by the soft-aperture effect of DFG, so as to produce the high quality THz vortex output in an eigenmode. In fact, the undesired higher-order radial modes are fully suppressed in the experimental THz vortex due to the soft-aperture DFG effect, as shown in Fig. 6(b). It is still difficult to further discuss the mode purity of the vortex output, for instance, OAM decomposition approach [67], owing to the absence of a commercial spatial light modulator (SLM) in the THz region.

It should also be noted that strong focusing of the vortex and Gaussian beams degrades the spatial quality of the THz vortex output. When the focused beam radii, $\omega_1$ and $\omega_2$, were ~60 µm (these values were smaller than the wavelength of the generated THz output), the resultant THz output showed a Gaussian profile, as shown in Fig. 7 (b).

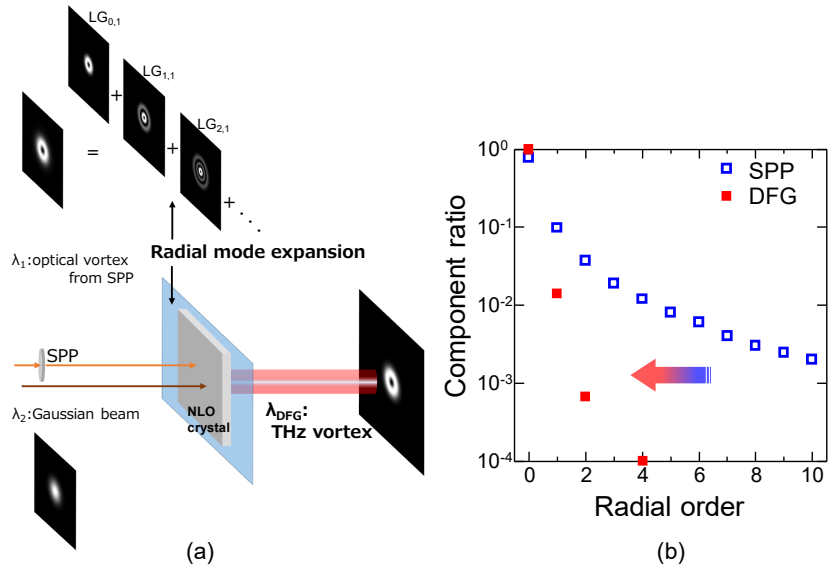

Fig. 5. (a) Soft-aperture effect via DFG to produce a high purity THz vortex output. (b) Simulated component ratios of the vortex outputs with ℓ=1 generated from the SPP and DFG as a function of the radial index, *p*. The undesired higher-order radial modes are suppressed in the vortex output generated from the DFG due to soft-aperture effects.

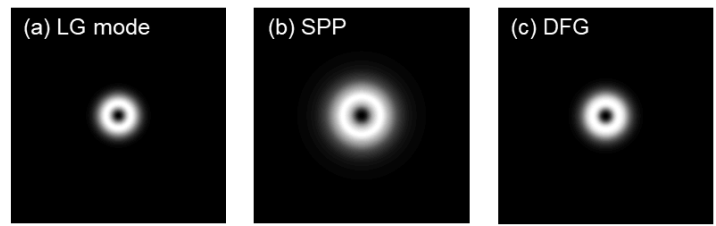

Fig.6 Simulated spatial forms of (a) a pure $LG_{0,1}$ mode, (b) optical vortex output produced by the SPP and (c) vortex mode generated by DFG.

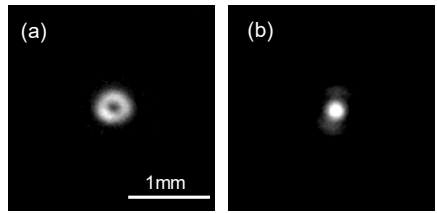

Fig. 7. Spatial distributions of the DFG output at 4 THz when the diameters of the pump beams were (a) ~170 µm and (b) less than 60 µm. [Note that (a) is the same as presented in Fig. 3(b)]

## 4. Conclusion

We have demonstrated, for the first time to the best of our knowledge, the generation of a widely tunable THz vortex output with high quality in an eigenmode by employing soft-aperture DFG of vortex and Gaussian outputs. The system enabled the production of a high-quality THz vortex mode with $\ell_{THz} = \pm 1$ within the frequency range of 2–6 THz. Such a tunable THz vortex source with high quality in an eigenmode will pave the way towards new advanced technologies, such as 2-dimensional identification of the crystallinity and polymorphism of organic crystals with high spatial resolution beyond the diffraction limit. This proposed method based on DFG should be easily extended to freely synthesize any THz wavefront, such as higher-order LG, Bessel, and Airy wavefronts, by freely modulating the wavefront of the pump beam using an SLM [68,69].

Our current system enables us to provide only one polarized state, however, it should be potentially extended to generate two orthogonally polarized eigen states by employing additional polarization elements or parallel DFGs with different polarization states. Thus, the generation of Poincaré modes, for instance a vector vortex mode, superimposed by two orthogonally polarized eigen states, will be possible as a future work.

## Appendix

*THz vortex output with topological charge $\ell_{THz}$ = ±2.*

There is no theoretical limitation for the topological charge $\ell$ of generated vortex modes. In fact, high quality second-order THz vortex outputs with a single ring without any undesired multiple outer rings, were also obtained (Figs. 8 (a), (b)). Their topological charges were also assigned to be $\ell_{THz} = \pm 2$ by the focusing method with the inclined lens (Figs. 8 (c), (d)). These indicate that our proposed approach allows us to generate the higher-order THz vortex modes with high quality.

In general, the OPA1 output with a larger $\ell$, created by using the SPP, exhibits a larger focal spot on the DAST crystal, thereby resulting in less spatial overlap with the OPA 2 output with the Gaussian profile. It is noteworthy that the DFG efficiency is expected to decrease as the increase of topological charge of $\ell$. In fact, the optical-optical efficiency of the second-order vortex output was measured to be ~$2.3 \times 10^{-4}$ %.

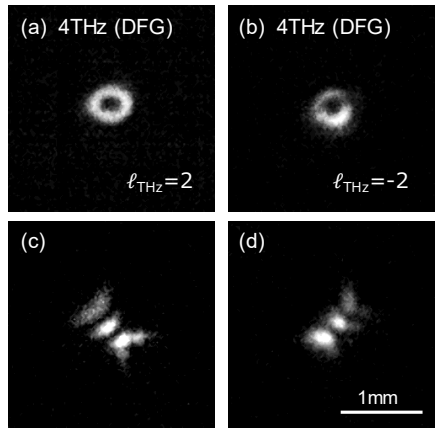

Fig. 8 (a,b) Far-fields of the 4 THz vortex outputs with $\ell_{THz} = \pm 2$ generated from the soft-aperture DFG. (c, g) Spatial forms of 4 THz vortex outputs focused by an inclined lens.

**Funding**

This work was funded by Kakenhi Grants-in-Aid (Nos. JP 19K05299, JP 18H03884 and JP 16H06507 ("Nano-Material Optical-Manipulation")) from the Japan Society for the Promotion of Science (JSPS).

**Disclosures**

The authors declare no conflicts of interest.